\documentclass[a4paper,12pt]{article}

\usepackage{amsmath,amssymb,indentfirst,epsfig,multirow,amsthm,color,array,amsfonts,graphicx,fullpage,hyperref,bm}
\usepackage[authoryear]{natbib}
\usepackage{amssymb,amsthm,epsfig, upgreek}
\usepackage{bm}
\usepackage{longtable}
\usepackage{threeparttable}

\usepackage[latin1]{inputenc}
\usepackage[figuresright]{rotating}
\usepackage{url,booktabs}

\usepackage{subfigure}

\begin{document}
\title{Beta and Kumaraswamy distributions as non-nested hypotheses in the modeling of continuous bounded data}

\author{Rodrigo B. Silva$^{\star,}$\footnote{Corresponding author. E-mail: \href{mailto:rodrigobs29@gmail.com}{rodrigobs29@gmail.com; rodrigo@de.ufpb.br}}\,\,\, and \,\,\,Wagner Barreto-Souza$^{\ddag,}$\footnote{E-mail: \href{mailto:wagnerbs@est.ufmg.br}{wagnerbs85@gmail.com; wagnerbs@est.ufmg.br}}\\\\
\centerline{\small{\it $^{\star}$Departamento de Estat\' \i stica, Universidade Federal da Para\'iba,}}\\
\centerline{\small{\it Cidade Universit\' aria - 58051-900, Jo\~ao Pessoa-PB, Brazil.}}\\
\centerline{\small{\it $^{\ddag}$Departamento de Estat\' \i stica, Universidade Federal de Minas Gerais,}}\\
\centerline{\small{\it Pampulha - 31270-901, Belo Horizonte-MG, Brazil.}}\\
}
\date{}
\sloppy
\maketitle

\begin{abstract}
Nowadays, beta and Kumaraswamy distributions are the most popular models to fit continuous bounded data. These models present some characteristics in common and to select one of them in a practical situation can be of great interest. With this in mind, in this paper we propose a method of selection between the beta and Kumaraswamy distributions. We use the logarithm of the likelihood ratio statistic (denoted by $T_n$, where $n$ is the sample size) and obtain its asymptotic distribution under the hypotheses $H_{\mathcal B}$ and $H_{\mathcal K}$, where $H_{\mathcal B}$ ($H_{\mathcal K}$) denotes that the data come from the beta (Kumaraswamy) distribution. Since both models has the same number of parameters, based on the Akaike criterion, we choose the model that has the greater log-likelihood value. We here propose to use the probability of correct selection (given by $P(T_n>0)$ or $P(T_n<0)$ depending on the null hypothesis) instead of only to observe the maximized log-likelihood values. We obtain an approximation for the probability of correct selection under the hypotheses $H_{\mathcal B}$ and $H_{\mathcal K}$ and select the model that maximizes it. A simulation study is presented in order to evaluate the accuracy of the approximated probabilities of correct selection. We illustrate our method of selection in two applications to real data sets involving proportions.\\

\end{abstract}

\section{Introduction}
\label{intro}

\noindent In Statistics, the beta distribution is a well-known and established model to fit continuous bounded data. A random variable $X$ following a beta distribution with shape parameters $a$ and $b$ has density function given by
\begin{equation}\label{pdfbeta}
f_{\mathcal B}(x; a, b) = \frac{1}{B(a,b)}x^{a-1}(1-x)^{b-1}, \quad 0<x<1, 
\end{equation}
where $B(a, b) = \int_0^1x^{a-1}(1-x)^{b-1}dx$ is the beta function; we denote $X\sim\mathcal B(a,b)$. We restrict our attention to the interval $(0,1)$ since a beta distribution on an interval $(c,d)$ (with $c<d$) is obtained by the simple linear transformation $(d-c)X+c$.

As an alternative to the beta distribution, \citet{kw1980} introduced a two-parameter distribution on $(0, 1)$, the so-called Kumaraswamy distribution. A random variable $Y$ following a Kumaraswamy distribution has density given by
\begin{equation}\label{pdfkw}
f_{\mathcal K}(y; \alpha, \beta) = \alpha \beta y^{\alpha-1}(1-y^\alpha)^{\beta-1}, \quad 0<y<1, 
\end{equation}
where $\alpha>0$ and $\beta>0$ are shape parameters. We denote $Y\sim\mathcal K(\alpha,\beta)$. Similarly as discussed above, we also restrict our attention to the Kumaraswamy distribution on the interval $(0,1)$. The Kumaraswamy distribution was initially proposed for applications in hydrology. Since then, it has been frequently used in several areas of Statistics in the last years. For instance, see the most recent papers by \citet{nad2008}, \citet{jones2009}, \citet{lem2011}, \citet{mit2013}, \citet{mitbae2013} and the references contained therein. One factor for this increased interesting on the Kumaraswamy distribution is due to its simple mathematical form of the distribution function, in constrast with the beta distribution. On the other hand, the ordinary moments of the beta distribution are obtained explicitly, while those of the Kumaraswamy distribution depend on the gamma function. There exist several advantages (and evidently disadvantages) of the Kumaraswamy distribution over the beta distribution. We recommend the paper by \citet{jones2009} to the readers interested in a detailed comparison between the beta and Kumaraswamy distributions.

Nowadays, the beta and Kumaraswamy distributions are the most popular models to fit continuous bounded data. Further, these models have many features in common and in a practical situation one question of interest is how to select the most adequate model (between the beta and Kumaraswamy distributions)  to fit a certain continuous bounded data set. To the best of our knowledge, it does not exist a way to discriminate the beta and Kumaraswamy models. In practical situations, the Akaike criteria has been used to do this, but this relies only on checking what is the model with great value of the maximized log-likelihood (since both have the same number of parameters). 


Our chief goal in this paper is to propose a selection criterion between the beta and Kumaraswamy distributions based on the asymptotic distribution of the likelihood ratio statistic proposed by \citet{cox1961,cox1962}. With this, we obtain the probability of correct selection under the hypotheses that the data comes from the beta or Kumaraswamy distributions and select the model that maximizes it.

In a pioneering work, \citet{cox1961,cox1962} proposed a way to discriminate non-nested families of hypotheses. The test statistic is the logarithm of the ratio of the maximized log-likelihoods under both null and alternative hypotheses. This statistic is compared with its expected value under the null hypothesis. Small deviations of the expected mean imply evidences in favor of the null hypothesis, while large deviations indicate evidences against. In a non-rigorous way, \citet{cox1962} showed that the normalized logarithm of the ratio of the maximized log-likelihoods is asymptotically normal distributed. Regularity conditions and a rigorous proof of the asymptotic normality of the Cox's test statistic was provided by \citet{white1982}.

The major part of the works dealing on this subject lies in discriminating between two non-nested lifetime distributions. For instance, see the papers by \citet{baieng1980}, \citet{feaneb1991}, \citet{gupkun2004}, \citet{kunetal2005}, \citet{deykun2012} and \citet{barsil2013}. References about discrimination between separate families of hypotheses are widespread and we recommend the reader to see references contained in the above papers.

This paper is outlined as follows. In Section~\ref{Likelihoodratiotest} we present the test statistic to discriminate beta and Kumaraswamy models and obtain its asymptotic distribution under two null hypotheses (that are, data come from the beta or Kumaraswamy distributions). In Section \ref{selectionmodel} we present our selection criterion based on the results given in the previous section. The minimum sample size required to discriminate beta and Kumaraswamy distributions when the probability of correct selection is beforehand is provided in Section~\ref{mss}. Simulation issues and two applications to real data sets involving proportions are presented in Sections~\ref{simulation} and~\ref{applications}, respectively. 

\section{Asymptotic distribution of the likelihood ratio statistic}
\label{Likelihoodratiotest}

Let $X_1, \ldots, X_n$ be a sequence of independent and identically distributed (iid) random variables, with observed values $x_1,\ldots, x_n$, either from a $\mathcal{B}(a, b)$ distribution or $\mathcal{K}(\alpha, \beta)$ distribution, with densities given by (\ref{pdfbeta}) and (\ref{pdfkw}), respectively. These hypotheses are denoted by
\begin{eqnarray*}
H_{\mathcal{B}}: \{X_i\}_{i=1}^n\sim \mathcal{B}(a, b) \,\,\, \mathrm{and}\,\,\, H_{\mathcal{K}}: \{X_i\}_{i=1}^n\sim \mathcal{K}(\alpha, \beta). 
\end{eqnarray*}

The log-likelihood function associated to the beta distribution is given by 
\begin{eqnarray*}\label{loglikbeta}
\ell^{(n)}_{\mathcal{B}}(a, b) = -n\log B(a,b) + (a-1) \sum_{i=1}^{n}\log x_i + (b-1) \sum_{i=1}^{n}\log (1-x_i). 
\end{eqnarray*}

The maximum likelihood estimates (MLEs) $\widehat{a}_n$ and $\widehat{b}_n$ of $a$ and $b$, respectively, are obtained as solutions of the nonlinear equations 
\begin{eqnarray*}
\psi(\widehat{a}_n) - \psi(\widehat{a}_n+\widehat{b}_n) - \frac{1}{n}\sum_{i=1}^n \log x_i =0 \quad \mathrm{and} \quad \psi(\widehat{b}_n) - \psi(\widehat{a}_n+\widehat{b}_n) - \frac{1}{n}\sum_{i=1}^n \log (1-x_i) = 0.  
\end{eqnarray*}
where $\psi(\cdot) $ is the digamma\footnote[1]{We denote generally the polygamma function by $\psi^{(m)}(\cdot), \,m=0, 1, \ldots$, where $\psi^{(m)}(x) = (d^{m+1}/dx^{m+1})\log\Gamma(x), \,x>0$.} function. On the other hand, the log-likelihood function corresponding to the Kumaraswamy distribution is
\begin{eqnarray*}\label{loglikkw}
\ell^{(n)}_{\mathcal{K}}(\alpha, \beta) = n\log \alpha + n \log \beta + (\alpha-1) \sum_{i=1}^{n}\log x_i + (\beta-1) \sum_{i=1}^{n}\log (1-x_i^\alpha). 
\end{eqnarray*}
The MLEs $\widehat\alpha_n$ and $\widehat\beta_n$ of $\alpha$ and $\beta$, respectively, are the solution of the nonlinear system of equations
\begin{eqnarray*}
\widehat{\beta}_n = -\frac{n}{\sum\limits_{i=1}^{n} \log(1-x_i^{\widehat{\alpha}_n})} \quad \mathrm{and} \quad \frac{n}{\widehat{\alpha}_n} + \sum_{i=1}^{n} \log x_i - (\widehat{\beta}_n-1) \sum_{i=1}^{n} \frac{x_i^{\widehat{\alpha}_n}\log x_i}{1-x_i^{\widehat{\alpha}_n}} =0. 
\end{eqnarray*}

With the above results, we define our test statistic by
\begin{eqnarray}\label{Tstatistic}
T_n=\log\left(\frac{\prod_{i=1}^nf_{\mathcal{B}}(x_i;\widehat{a}_n,\widehat{b}_n)}{\prod_{i=1}^nf_{\mathcal{K}}(x_i;\widehat{\alpha}_n,\widehat{\beta}_n)}\right)=\ell_{\mathcal{B}}(\widehat{a}_n,\widehat{b}_n)-\ell_{\mathcal{K}}(\widehat\alpha_n,\widehat\beta_n),
\end{eqnarray}
where $(\widehat{a}_n,\widehat{b}_n)$ and $(\widehat{\beta}_n,\widehat{\alpha}_n)$ are the MLEs of $(a,\, b)$ and $(\beta,\alpha)$, respectively. In words, our test statistic is the difference between the maximized log-likelihoods. Since both models have the same number of parameters, this corresponds to the Akaike statistic (\citet{aka1974}). More explicitly, the statistic $T_n$ can be expressed as
\begin{eqnarray*}
T_n &=& n\left[1-\log B(\widehat a_n, \widehat b_n) - \log (\widehat \alpha_n \widehat \beta_n)\right] + n(\widehat a_n - \widehat \alpha_n)\left[\psi(\widehat a_n) - \psi(\widehat a_n + \widehat b_n)\right]\\
&+&n(\widehat b_n - 1)\left[\psi(\widehat b_n) - \psi(\widehat a_n + \widehat b_n)\right]+ \sum_{i=1}^{n}\log(1-x^{\widehat \alpha_n})
\end{eqnarray*}

In practical situations, based on the Akaike criterion, the following selection criterion is commonly adopted: we choose the beta distribution if $T_n>0$, otherwise we choose the Kumaraswamy distribution. We here adopt a different selection criterion, which is based on the asymptotic distribution of a normalized version of $T_n$ under the hypotheses $\mathcal{H}_{\mathcal B}$ and $\mathcal{H}_{\mathcal K }$. This criterion will be present in the next section. Now we concentrate our attention to find the asymptotic distribution of the test statistic. We now define some function which will appear along the paper. For $x,y,z>0$, define the real functions
\begin{eqnarray}\label{F}
\mathcal F(x,y,z)=\sum_{k=1}^\infty k^{-1}B(x+kz,y),
\end{eqnarray}
\begin{eqnarray}\label{G}
\mathcal G(x,y,z)=\sum_{k=1}^\infty\{\psi(x+kz)-\psi(x+y+kz)\}B(x+kz,y),
\end{eqnarray}
\begin{eqnarray}\label{M}
\mathcal M(x,y,z)=\sum_{k=1}^\infty k^{-1}\{\psi(x+kz)-\psi(x+y+kz)\}B(x+kz,y),
\end{eqnarray}
\begin{eqnarray}\label{V}
\mathcal V(x,y,z)=\sum_{k=1}^\infty k^{-1}\{\psi(y)-\psi(x+y+kz)\}B(x+kz,y)
\end{eqnarray}
and
\begin{eqnarray}\label{W}
\mathcal W(x,y,z)=\sum_{k=1}^\infty(-1)^k \frac{\{[\psi(1)-\psi((x+k)z^{-1}+1)]^2+\psi'(1)-\psi'((x+k)z^{-1}+1)\}}{\Gamma(y-k)k!(x+k)},
\end{eqnarray}
where $\psi'(\cdot)$ is the first derivative of the digamma function $\psi(\cdot)$. 

\subsection{Beta distribution as the null hypothesis}\label{case1} 

In this subsection we present the asymptotic distribution of $T_n$ under the hypothesis $H_\mathcal{B}$. The alternative hypothesis is $H_\mathcal{K}$.
Suppose that the random variables $X_1,\ldots,X_n$ come from the $\mathcal{B}(a, b)$ distribution. For any Borel measurable function $h(\cdot)$, the underscript $\mathcal B$ in $E_{\mathcal B}(h(X_1))$ means that the expectation is taken with respect to the beta distribution with density given by (\ref{pdfbeta}). More explicitly, we have $E_{\mathcal B}(h(X_1))=\int_0^1h(x)f_{\mathcal B}(x;a,b)dx$.\\

 Under the hypothesis $\mathcal H_{\mathcal B}$, as $n\rightarrow\infty$ we have that\\

\noindent(i) $\widehat{a}_n\longrightarrow a$ and $\widehat{b}_n\longrightarrow b$ almost surely, where 
$$E_{\mathcal B}(\log f_{\mathcal B}(X; a, b)=\max_{\bar{a}, \bar{b}}E_{\mathcal B}(\log f_{\mathcal B}(X;\bar{a}, \bar{b}));$$

\noindent(ii) $\widehat{\alpha}_n\longrightarrow\widetilde{\alpha}$ and $\widehat{\beta}_n\longrightarrow\widetilde{\beta}$ almost surely, where 
$$E_{\mathcal B}(\log f_{\mathcal K}(X;\widetilde{\alpha},\widetilde{\beta}))=\max_{\alpha,\beta}E_{\mathcal B}(\log f_{\mathcal K}(X;\alpha,\beta)).$$

The quasi-maximum likelihood estimators $\widetilde{\alpha}$ and $\widetilde{\beta}$ are functions of $a$ and $b$, which is not explicited in order to simplify the notation.
The above convergences follow from the results stated and proved by \citet{white1982b}.

We now discuss how to obtain $\widetilde{\alpha}$ and $\widetilde{\beta}$. Define $\Lambda_{\mathcal{B}}(\alpha, \beta) = \mathrm{E}_{\mathcal{B}}(\log f_{\mathcal{K}}(X; \alpha,\beta))$. We have that
\begin{eqnarray*}
\Lambda_{\mathcal{B}}(\alpha, \beta) = \log \alpha + \log \beta + (\alpha-1)(\psi(a) - \psi(a+b)) - \frac{(\beta-1)}{B(a,b)}\mathcal F(a,b,\alpha).
\end{eqnarray*}

With this, we have that $\widetilde{\alpha}$ and $\widetilde{\beta}$ are obtained as the solution of the system of nonlinear equations $(\partial \Lambda_{\mathcal{B}}(\alpha, \beta) /\partial\alpha,\partial \Lambda_{\mathcal{B}}(\alpha, \beta) /\partial\beta)^\top=(0,0)^\top$, that is
\begin{equation*}
\widetilde{\beta} = \frac{B(a, b)}{\mathcal F(a,b,\widetilde\alpha)}\quad \mbox{and}\quad \frac{1}{\widetilde{\alpha}} + \psi(a) - \psi(a+b) - \frac{(\widetilde{\beta}-1)}{B(a,b)}\mathcal G(a,b,\widetilde\alpha) =0.
\end{equation*}

Now, in order to present the asymptotic distribution of the test statistic $T_n$ under $\mathcal H_{\mathcal B}$, we need to compute the mean and variance of the random variable $\log f_{\mathcal B}(X; a, b)-\log f_{\mathcal K}(X;\widetilde\alpha,\widetilde\beta)$ (with $X\sim{\mathcal B}(a, b)$), which we will be denoted by $\mbox{AM}_{\mathcal B}(a,b)$ and $\mbox{AV}_{\mathcal B}(a,b)$, respectively.
 
An explicit expression for $\mbox{AM}_{\mathcal{B}}(a, b)$ is given by
\begin{eqnarray}\label{AMbe}\nonumber
\mbox{AM}_{\mathcal{B}}(a, b) &=&-\log \widetilde{\alpha} - \log \widetilde{\beta} - \log B(a, b) - (\widetilde{\alpha}- a)(\psi(a) - \psi(a+b))+\\
 && (b-1)(\psi(b) - \psi(a+b)) - \frac{(\widetilde{\beta}-1)}{B(a,b)}\mathcal F(a,b,\widetilde\alpha).
\end{eqnarray}
where the real function $\mathcal F(\cdot,\cdot,\cdot)$ was defined in (\ref{F}). The variance $\mbox{AV}_{\mathcal{B}}(a, b)$ is given by
\begin{eqnarray}\label{AVbe} \nonumber
\mbox{AV}_{\mathcal B}(a,b) &=& (\widetilde{\alpha}-a)^2 \mathrm{Var}_{\mathcal B}(\log X) + (\widetilde{\beta}-1)^2\mathrm{Var}_{\mathcal B}(\log (1-X^{\widetilde{\alpha}}))+\\ \nonumber
 && (b-1)^2\mathrm{Var}_{\mathcal B}(\log (1-X)) + 2(\widetilde{\beta}-1)(\widetilde{\alpha}-a)\mathrm{Cov}_{\mathcal B}(\log X, \log (1-X^{\widetilde{\alpha}}))-\\ \nonumber
&&2(b-1)(\widetilde{\alpha}-a)\, \mathrm{Cov}_{\mathcal B}(\log X, \log (1-X))-\\  
&& 2(\widetilde{\beta}-1)(b-1)\,\mathrm{Cov}_{\mathcal B}(\log (1-X), \log (1-X^{\widetilde{\alpha}})),
\end{eqnarray}
where the variances and covariances above can be expressed by 
\begin{eqnarray*}
&&\mathrm{Var}_{\mathcal B}(\log X) = \psi'(a) - \psi'(a+b), \quad \mathrm{Var}_{\mathcal B}(\log (1-X)) = \psi'(b) - \psi'(a+b)\\ 
&&\mathrm{Var}_{\mathcal B}(\log (1-X^{\widetilde{\alpha}})) = \frac{\Gamma(a+b)}{\Gamma(a)}\mathcal W(a,b,\widetilde\alpha)-\frac{1}{B(a,b)^2}\mathcal F(a,b,\widetilde\alpha)^2,\\
&&\mathrm{Cov}_{\mathcal B}(\log X, \log (1-X^{\widetilde{\alpha}})) = \frac{1}{B(a,b)}\left\lbrace \left[ \psi(a)-\psi(a+b)\right]\mathcal F(a,b,\widetilde\alpha)-\mathcal M(a,b,\widetilde\alpha)\right\rbrace ,\\
&&\mathrm{Cov}_{\mathcal B}(\log (1-X), \log (1-X^{\widetilde{\alpha}})) =  \frac{1}{B(a,b)}\left\lbrace \left[ \psi(a)-\psi(a+b)\right] \mathcal F(a,b,\widetilde\alpha)-\mathcal V(a,b,\widetilde\alpha)\right\rbrace ,\\
&&\mathrm{Cov}_{\mathcal B}(\log X, \log (1-X)) = -\psi'(a+b),
\end{eqnarray*}
with $\mathcal F(\cdot,\cdot,\cdot)$, $\mathcal M(\cdot,\cdot,\cdot)$, $\mathcal V(\cdot,\cdot,\cdot)$ and $\mathcal W(\cdot,\cdot,\cdot)$ as defined in (\ref{F}), (\ref{M}), (\ref{V}) and (\ref{W}), respectively.

Table \ref{numericbe} lists the values of $\mbox{AM}_{\mathcal{B}}(a, b)$, $\mbox{AV}_{\mathcal{B}}(a, b)$, $\widetilde{\alpha}$ and $\widetilde{\beta}$ for $b=2.5$ and some values of the parameter $a$.
\begin{table}[!htbp]
\centering
\caption{Values of $\mbox{AM}_\mathcal{B}(a, b)$, $\mbox{AV}_\mathcal{B}(a, b)$, $\widetilde{\alpha}$ and $\widetilde{\beta}$ for $b=2.5$ and some values of $a$.}
\renewcommand{\arraystretch}{1.3}
\scalebox{1}[0.87]{
	\begin{tabular}{ccccccccccccccccccccccccccc}
	\hline
$a$ &&&&& $\mbox{AM}_{\mathcal B}(a, b)$ &&&&& $\mbox{AV}_{\mathcal B}(a, b)$ &&&&& $\widetilde{\alpha}$ &&&&& $\widetilde{\beta}$\\	
\hline
0.2&&&&&	0.003827&&&&&	0.008466&&&&&	0.2242&&&&&	1.5522\\[0.05in]
0.5&&&&&	0.000644&&&&&	0.002422&&&&&	0.5383&&&&&	1.8378\\[0.05in]
0.7&&&&&	0.000072&&&&&	0.001804&&&&&	0.7616&&&&&	1.9262\\[0.05in]
1.2&&&&&	0.000065&&&&&	0.000975&&&&&	1.1734&&&&&	2.0299\\[0.05in]
1.5&&&&&	0.000033&&&&&	0.001165&&&&&	1.4270&&&&&	2.0591\\[0.05in]
2.0&&&&&	0.000192&&&&&	0.001470&&&&&	1.8388&&&&&	2.0866\\
	\hline
	\end{tabular}}\label{numericbe}
\end{table}

We now present the asymptotic distribution of $n^{-1/2}(T_n-E_{\mathcal B}(T_n))$. Define $\widetilde{T}^{\mathcal B}_n=\ell^{(n)}_{\mathcal{B}}(a, b)-\ell^{(n)}_{\mathcal{K}}(\widetilde\alpha,\widetilde\beta)$.

Under the null hypothesis $\mathcal H_{\mathcal B}$,  we have that
\begin{eqnarray}\label{theorem-be}
n^{-1/2}(T_n- \mbox{E}_{\mathcal B}(T_n))\sim n^{-1/2}(\widetilde T^{\mathcal B}_n-n \mbox{AM}_{\mathcal B}(a, b))
 \stackrel{d}{\longrightarrow} N(0, \mbox{AV}_{\mathcal B}(a, b)),
\end{eqnarray}
as $n\rightarrow\infty$, where $\mbox{AM}_{\mathcal B}(a, b)$ and $\mbox{AV}_{\mathcal B}(a, b)$ are given by (\ref{AMbe}) and (\ref{AVbe}), respectively, and ``$\sim$'' denotes ``asymptotically equivalent''.

We now justify that the above result is in fact true. From the Central Limit Theorem, it follows that $n^{-1/2}(\widetilde T^{\mathcal B}_n-n \mbox{AM}_{\mathcal B}(a, b))\stackrel{d}{\longrightarrow}\mathrm{N}(0, \mbox{AV}_{\mathcal B}(a, b))$ as $n\rightarrow\infty$. Therefore, the major work in proving (\ref{theorem-be}) lies in showing the asymptotic equivalence between $n^{-1/2}(T_n- \mbox{E}_{\mathcal B}(T_n))$ and $n^{-1/2}(\widetilde T^{\mathcal B}_n-n \mbox{AM}_{\mathcal B}(a, b))$.
This follows from an adaptation of the results given in \citet{white1982}. This adaptation is made in \citet{barsil2013} for the discriminating between the exponential-Poisson and gamma distributions. Following exactly as made there, the results here presented follows.

\subsection{Kumaraswamy distribution as the null hypothesis}\label{case2} 

We now suppose that $H_\mathcal K$ and $H_\mathcal B$ are the null and alternative hypotheses, respectively. Let $X_1,\ldots,X_n$ be iid random variables following a $\mathcal{K}(\alpha, \beta)$ distribution. Similarly as in the previous case, for any Borel measurable function $h(\cdot)$, the underscript $\mathcal K$ in $\mathrm{E}_{\mathcal{K}}(h(X_1))$ means that the expectation is taken with respect to the Kumaraswamy distribution with density given in (\ref{pdfkw}), that is,  $\mathrm{E}_{\mathcal{K}}(h(X_1))=\int_0^1h(x)f_{\mathcal K}(x;\alpha,\beta)dx$.\\

Under the hypothesis $\mathcal H_{\mathcal K}$, as $n\rightarrow\infty$ we have that\\

\noindent(i) $\widehat{\alpha}_n\longrightarrow\alpha$ and $\widehat{\beta}_n\longrightarrow\beta$ almost surely, where 
$$\mbox{E}_{\mathcal K}(\log f_{\mathcal K}(X; \alpha, \beta))=\max_{\bar{\alpha}, \bar{\beta}}\mbox{E}_{\mathcal K}(\log f_{\mathcal K}(X;\bar{\alpha}, \bar{\beta}));$$

\noindent(ii) $\widehat{a}_n\longrightarrow\widetilde{a}$ and $\widehat{b}_n\longrightarrow\widetilde{b}$ almost surely, where 
$$\mbox{E}_{\mathcal K}(\log f_{\mathcal B}(X;\widetilde{a},\widetilde{b}))=\max_{a, b}\mbox{E}_{\mathcal K}(\log f_{\mathcal B}(X; a, b)).$$

As before, we call attention of the reader that the quasi-maximum likelihood estimators $\widetilde{a}$ and $\widetilde{b}$ are functions of $\alpha$ and $\beta$, which is not explicited for brevity.

We now show how to obtain $\widetilde{a}$ and $\widetilde{b}$. Define $\Lambda_{\mathcal{K}}(a, b)=\mathrm{E}_{\mathcal{K}}(\log f_{\mathcal{B}}(X; a, b))$. We have that
\begin{eqnarray*}
\Lambda_{\mathcal{K}}(a, b) = -\log B(a, b) + \frac{(a-1)}{\alpha}(\psi(1) - \psi(\beta+1)) - \beta(b-1) \mathcal F(1,\beta,\alpha^{-1}).\\\
\end{eqnarray*}

Hence, $\widetilde a$ and $\widetilde b$ are obtained as solution of the system of nonlinear equations $(\partial \Lambda_{\mathcal{K}}(a, b)/\partial a, \Lambda_{\mathcal{K}}(a, b)/\partial b)^\top=(0,0)^\top$. These equations are given by
\begin{align*}
&\psi(\widetilde a+ \widetilde b) - \psi(\widetilde a) + \frac{1}{\alpha}\left[\psi(1) - \psi(\beta+1)\right] = 0\\ 
\intertext{and}
&\psi(\widetilde a+\widetilde b) - \psi(\widetilde b) - \beta \mathcal F(1,\beta,\alpha^{-1})=0.
\end{align*}

We now compute the mean and variance of the random variable $\log f_{\mathcal B}  (X; \widetilde a, \widetilde b)-\log f_{\mathcal K}(X; \alpha, \beta)$ (with $X\sim{\mathcal K}(\alpha, \beta)$), which we will denote by $\mbox{AM}_{\mathcal K}(\alpha,\beta)$ and $\mbox{AV}_{\mathcal K}(\alpha,\beta)$, respectively. As in the previous case, these results will be important to present the asymptotic distribution of the test statistic $T_n$ under $\mathcal{H}_{\mathcal K}$. After some algebra, it can be shown that these quantities can be expressed by
\begin{eqnarray}\label{AMkw}\nonumber
\mbox{AM}_{\mathcal{K}}(\alpha, \beta) &=& -\log\alpha - \log\beta - \frac{(\alpha - \widetilde a)}{\alpha}(\psi(1) - \psi(\beta+1)) - \log B(\widetilde a, \widetilde b)+\\ 
&&\frac{\beta -1}{ \beta} - \beta(\widetilde b -1)\mathcal F(1,\beta,\alpha^{-1})
\end{eqnarray}
and 
\begin{eqnarray}\label{AVkw} \nonumber
\mbox{AV}_{\mathcal{K}}(\alpha, \beta) &=& (\alpha-\widetilde a)^2\mbox{Var}_{\mathcal K}(\log X) + (\beta -1)^2\mbox{Var}_{\mathcal K}(\log (1-X^\alpha))+ \\ \nonumber
&& (\widetilde b - 1)^2 \mbox{Var}_{\mathcal K}(\log(1-X)) + 2(\alpha-\widetilde a)(\beta-1)\mbox{Cov}_{\mathcal K}(\log X, \log (1-X^\alpha))-\\ \nonumber
&& 2(\alpha-\widetilde a)(\widetilde b-1)\mbox{Cov}_{\mathcal K}(\log X, \log (1-X))-\\
 && 2(\widetilde b-1)(\beta-1)\mbox{Cov}_{\mathcal K}(\log(1-X), \log (1-X^\alpha)),
\end{eqnarray}
where the variances and covariances above can be expressed by
\begin{eqnarray*}
&&\mbox{Var}_{\mathcal K}\left(\log X\right) = \frac{1}{\alpha^2}(\psi'(1) - \psi'(\beta+1)),\quad \mbox{Var}_{\mathcal K}(\log(1-X^\alpha)) = \frac{1}{\beta^2}, \\
&&\mbox{Var}_{\mathcal K}(\log(1-X)) = \Gamma(\beta+1)\mathcal V(1,\beta,\alpha^{-1})-\beta^2\mathcal F(1,\beta,\alpha^{-1}),\\
&&\mbox{Cov}_{\mathcal K}(\log X, \log(1-X))=\frac{\beta}{\alpha}\left\lbrace \left[ \psi(1) - \psi(\beta+1)\right] \mathcal{F}(1,\beta,\alpha^{-1})-\mathcal M(1,\beta,\alpha^{-1})\right\rbrace ,\\
&&\mbox{Cov}_{\mathcal K}(\log(1-X), \log(1-X^\alpha))=-\beta\left[ \mathcal V(1,\beta,\alpha^{-1})+\beta\mathcal F(1,\beta,\alpha^{-1})\mathcal F(1,\beta,1)\right] ,\\
&&\mbox{Cov}_{\mathcal K}(\log X, \log(1-X^\alpha))=-\frac{\psi'(\beta+1)}{\alpha}.
\end{eqnarray*}

The real functions $\mathcal F(\cdot,\cdot,\cdot)$, $\mathcal M(\cdot,\cdot,\cdot)$ and $\mathcal V(\cdot,\cdot,\cdot)$ that appear above are defined in (\ref{F}), (\ref{M}) and (\ref{V}), respectively. Table \ref{numerickw} presents the values of $\mbox{AM}_{\mathcal{K}}(\alpha, \beta)$, $\mbox{AV}_{\mathcal{K}}(\alpha, \beta)$, $\widetilde{a}$ and $\widetilde{b}$ for $\beta=2.5$ and some values of the parameter $\alpha$.
\begin{table}[!htbp]
\centering
\caption{Values of $\mbox{AM}_{\mathcal{K}}(\alpha, \beta)$, $\mbox{AV}_{\mathcal{K}}(\alpha, \beta)$, $\widetilde{a}$ and $\widetilde{b}$ for $\beta=2.5$ and some values of $\alpha$.}
\renewcommand{\arraystretch}{1.3}
\scalebox{1}[0.87]{
\begin{tabular}{ccccccccccccccccccccccccccc}
\hline
$\alpha$ &&&&& $\mbox{AM}_{\mathcal{K}}(\alpha, \beta)$ &&&&& $\mbox{AV}_{\mathcal{\mathcal K}}(\alpha, \beta)$ &&&&& $\widetilde{a}$ &&&&& $\widetilde{b}$\\
\hline
0.2&&&&&	$-$0.011825&&&&&	0.746237&&&&&	0.1626&&&&&	3.0761\\[0.05in]
0.5&&&&&	$-$0.001315&&&&&	0.071849&&&&&	0.4549&&&&&	2.2410\\[0.05in]
0.7&&&&&	$-$0.000259&&&&&	0.014987&&&&&	0.6667&&&&&	2.0968\\[0.05in]
1.2&&&&&	$-$0.000037&&&&&	0.002621&&&&&	1.2292&&&&&	1.9668\\[0.05in]
1.5&&&&&	$-$0.000143&&&&&	0.010834&&&&&	1.5801&&&&&	1.9372\\[0.05in]
2.0&&&&&	$-$0.000294&&&&&	0.025130&&&&&	2.1773&&&&&	1.9122\\
\hline
\end{tabular}}
\label{numerickw}
\end{table}

Define now the quantity $\widetilde{T}^{\mathcal K}_n=\ell^{(n)}_{\mathcal{B}}(\widetilde{a},\widetilde{b})-\ell^{(n)}_{\mathcal{K}}(\alpha,\beta)$.
Under the hypothesis $\mathcal{H}_{\mathcal K}$, we have that
\begin{eqnarray}\label{theorem-kw}
n^{-1/2}(T_n-\mbox{E}_{\mathcal K}(T_n))\sim n^{-1/2}(\widetilde{T}^{\mathcal K}_n-n\mbox{AM}_{\mathcal K}(\alpha, \beta))
\stackrel{d}{\longrightarrow}N(0, \mbox{AV}_{\mathcal K}(\alpha, \beta)),
\end{eqnarray}
as $n\rightarrow\infty$, where $\mbox{AM}_{\mathcal K}(\alpha, \beta)$ and $\mbox{AV}_{\mathcal K}(\alpha, \beta)$ are given by (\ref{AMkw}) and (\ref{AVkw}), respectively. As before, ``$\sim$'' denotes ``asymptotically equivalent''.
The justification of the validality of the convergence given in (\ref{theorem-kw}) is exactly the same of the justification of the result (\ref{theorem-be}).

\section{Selection criterion}
\label{selectionmodel}

With the results presented in the previous section, we are ready to give our selection criterion. For this, let us first to present asymptotic forms for the probabilities of correct selection (in short PCS) $\mbox{PCS}_\mathcal B(a,b)\equiv P(T_n>0)$ and $\mbox{PCS}_\mathcal K(\alpha,\beta)\equiv P(T_n<0)$ under the hypotheses $H_\mathcal B$ and $H_\mathcal K$, respectively.

Assume that the null and alternative hypotheses are $H_\mathcal B$ and $H_\mathcal K$, respectively. From the result (\ref{theorem-be}), we have that ${\rm PCS_{\mathcal{B}}}(a, b)$ may be approximated by 
\begin{eqnarray}\label{pcsbe}
{\rm PCS_{\mathcal{B}}}(a, b)\approx \Phi\left(-\frac{\sqrt{n}{\mbox{AM}_{\mathcal{B}}}(a, b)}{\sqrt{\mbox{AV}_{\mathcal{B}}(a, b)}}\right),
\end{eqnarray}
where $\Phi(\cdot)$ is the distribution function of the standard normal distribution and $\mbox{AM}_{\mathcal{B}}(a, b)$ and $\mbox{AV}_{\mathcal{B}}(a, b)$ are given in (\ref{AMbe}) and (\ref{AVbe}), respectively.

Now consider the null and alternative hypotheses are $H_\mathcal B$ and $H_\mathcal K$, respectively. Based on the convergence in distribution given in (\ref{theorem-kw}), we have that ${\rm PCS_{\mathcal{K}}}(\alpha, \beta)$ may be approximated by 
\begin{eqnarray}\label{pcskw}
{\rm PCS_{\mathcal{K}}}(\alpha, \beta)\approx \Phi\left(\frac{\sqrt{n}{\mbox{AM}_{\mathcal{K}}}(\alpha, \beta)}{\sqrt{\mbox{AV}_{\mathcal{K}}(\alpha, \beta)}}\right),
\end{eqnarray}
where $\Phi(\cdot)$ is the distribution function of the standard normal distribution and $\mbox{AM}_{\mathcal{K}}(\alpha, \beta)$ and $\mbox{AV}_{\mathcal{K}}(\alpha, \beta)$ are given in (\ref{AMkw}) and (\ref{AVkw}), respectively. 

The probabilities of correct selection (\ref{pcsbe}) and (\ref{pcskw}) depend on the parameters. In practice, we replace the parameters by their maximum likelihood estimators. With this, we define our selection criterion as follows:

\begin{itemize}
\item If ${\rm PCS_{\mathcal{B}}}(\widehat a, \widehat b)>{\rm PCS_{\mathcal{K}}}(\widehat\alpha, \widehat\beta)$, choose the beta distribution, otherwise select the Kumaraswamy distribution, where $(\widehat a, \widehat b)$ and $(\widehat\alpha, \widehat\beta)$ are respectively the maximum likelihood estimators of $(a,b)$ and $(\alpha,\beta)$ given in the previous section.
\end{itemize}

The above selection criterion is alternatively equivalent to the following one:
\begin{itemize}
\item If ${\mbox{AM}_{\mathcal{K}}}(\widehat\alpha, \widehat\beta)\sqrt{\mbox{AV}_{\mathcal{B}}(\widehat a, \widehat b)}<-{\mbox{AM}_{\mathcal{B}}}(\widehat a,\widehat b)\sqrt{\mbox{AV}_{\mathcal{K}}(\widehat\alpha,\widehat\beta)}$, choose the beta distribution, otherwise select the Kumaraswamy distribution.
\end{itemize}

\section{Distances and minimum sample size}\label{mss}

We now propose a method to determine the minimum sample size required in order to discriminate between the beta and Kumaraswamy distributions for a specified PCS and a given tolerance level, which is defined in terms of some distance to measure the closeness between the beta and Kumaraswamy distributions. 

There are several ways to measure the closeness or the distance between two probability distributions. The most common measures are the Kolmogorov-Smirnov ($\mathcal{KS}$) and Hellinger ($\mathcal{H}$) distances and we will use both in this paper. 

Let $f$ and $g$ (with same support $\Omega$) be two absolutely continuous density functions with distribution functions $F(x)$ and $G(x)$, respectively. The Kolmogorov-Smirnov distance between $F$ and $G$ is given by
\begin{eqnarray*}
\mathcal{KS}(F,G)=\sup_{x\in\Omega}|F(x)-G(x)|.
\end{eqnarray*}
 The Hellinger distance between $f$ and $g$ is defined by
\begin{eqnarray*}
\mathcal{H}(f,g)=\frac{1}{2}\int_{\Omega}\left(\sqrt{f(x)}-\sqrt{g(x)}\right)^2dx=1-\int_{\Omega}\sqrt{f(x)g(x)}dx.
\end{eqnarray*}

It is not possible to find an explicit expression for the Kolmogorov-Smirnov distance in our case. On the other hand, we find an explicit expression for the Hellinger distance between beta and Kumaraswamy distributions, that is
\begin{eqnarray*}\label{hellingerbekw}
\mathcal{H}(f_{\mathcal{B}},f_{\mathcal{K}})=1-\left(\frac{\alpha\beta}{B(a, b)}\right)^{1/2}\sum_{k=0}^\infty (-1)^k  \binom{\frac{1}{2}(\beta-1)}{k} B\left(\frac{1}{2}\left[a+(2k+1)\alpha\right], \frac{1}{2}\left(b+1\right)\right)
\end{eqnarray*}

The above expression can be obtained by using the binomial expansion in $(1-x^\alpha)^{(\beta-1)/2}$ in $\sqrt{f_{\mathcal{B}}(x)f_{\mathcal{K}}(x)}$ and hence using the Dominate Convergence Theorem. 

For small distances between two probability distributions, it is expected that the minimum sample size required to discriminate them be large. Otherwise, a small or moderate sample size is sufficient to discriminate the models. We assume that the user will specify beforehand the PCS and the tolerance level in terms of the distance  between the beta and Kumaraswamy distributions. When a tolerance level is specified (by means of some distance), the two distribution functions are not considered to be significantly different if their distance does not exceed the tolerance level. PCS and tolerance level play a similar role that the power and Type-I error in the corresponding testing of hypotheses problem.

Based on PCS and tolerance level we can determine the minimum sample size required to discriminate between the beta and Kumaraswamy distributions. The tolerance level here is defined for the $\mathcal{KS}$ and $\mathcal{H}$ distances. We are now interested in finding the required sample size $n$ such that PCS achieves a certain protection level $p$ for a stated tolerance level $D$. 

We explain the procedure under the null hypothesis $H_{\mathcal{B}}$. The procedure under $H_{\mathcal{K}}$ follows in a similar way and therefore is omitted.

To determine the sample size needed to achieve at least a protection level $p$, we equate $\mbox{PCS}_\mathcal B(a,b)=p$. Hence, using the asymptotic result given in (\ref{pcsbe}) we get 
$$\Phi\left(-\frac{\sqrt{n}AM_{\mathcal{B}}(a, b)}{\sqrt{AV_{\mathcal{B}}(a, b)}}\right)=p.$$

By solving for $n$ we obtain
\begin{eqnarray}\label{nbe}
n=\left[ \frac{z^2_{p}AV_{\mathcal{B}}(a, b)}{AM^2_{\mathcal{B}}(a, b)}\right] ,
\end{eqnarray}
where $z_{p}$ is the 100$p$ percentile point of the standard normal distribution and $[z]$ denotes the smallest integer $y$ such that $y>z$, for $z\in\mathbb{R}$. Similarly, under the null hypothesis $\mathcal{H}_{\mathcal{B}}$ and using the result (\ref{pcskw}) we need
\begin{eqnarray}\label{nkw}
n=\left[\frac{z^2_{p}AV_{\mathcal{K}}(\alpha, \beta)}{AM^2_{\mathcal{K}}(\alpha, \beta)}\right] ,
\end{eqnarray}
to choose the Kumaraswamy distribution with ${\rm PCS}$ equal to $p$. Values of (\ref{nbe}) for some values of $a$, corresponding to $b=3$ and $p = 0.6, 0.7,0.8$, are given in Table \ref{ksbe}. Table \ref{kskw} lists some values of (\ref{nkw}) for given values of $\alpha$, with $\beta =2$ and $p = 0.6, 0.7,0.8$. In these tables, values of the $\mathcal{KS}$ and $\mathcal{H}$ distances are also presented.

\begin{table}[!htbp]
\centering
\caption{Values of $n$ and the $\mathcal{H}$ and $\mathcal{KS}$ distances between $\mathcal{B}(a,b)$ and $\mathcal{K}(\widetilde{\alpha},\widetilde{\beta})$ distributions for $b=3$ and some values of $a$.}\label{ksbe}
\scalebox{0.98}[0.90]{
\renewcommand{\arraystretch}{1.3}
\begin{tabular}{lllcccccccccccccccccccc}
\hline
$a \rightarrow$	&&& 0.2    &&& 0.5    &&&  1.5  &&&	2.0    &&&	3.0   &&&	5.0\\
\hline
$n \,(p=0.6)$	&&& 14    	&&& 75     &&&	380  	&&&	161    &&&	89    &&& 64\\
$n \,(p=0.7)$	&&& 60    	&&& 323    &&&	1630  &&&	692    &&&	384   &&& 275\\
$n \,(p=0.8)$	&&& 159    	&&& 859    &&&	4651  &&&	1783   &&&	989   &&& 708\\
$\mathcal{H}$ &&& 0.0022 	&&& 0.0004 &&& 0.0001 &&& 0.0002 &&& 0.0004 &&& 0.0005\\
$\mathcal{KS}$&&& 0.0104 	&&& 0.0000 &&& 0.0000 &&& 0.0000 &&& 0.0000 &&& 0.0110\\
\hline
\end{tabular}}
\end{table}

\begin{table}[!htbp]
\centering
\caption{Values of $n$ and the $\mathcal{H}$ and $\mathcal{KS}$ distances between $\mathcal{B}(\widetilde a,\widetilde b)$ and $\mathcal{K}(\alpha,\beta)$ distributions for $\beta=0.3$ and some values of $\alpha$.}\label{kskw}
\scalebox{0.98}[0.90]{
\renewcommand{\arraystretch}{1.3}
\begin{tabular}{lllcccccccccccccccccccc}
\hline
$\alpha \rightarrow$	&&&0.2&&&	0.5&&&	1.5&&&	2.0&&&	3.0&&&	5.0\\
\hline
$n \,(p=0.6)$	&&& 12    &&& 98     &&&	907  &&&	443     &&&	287    &&& 233\\
$n \,(p=0.7)$	&&& 47    &&& 417    &&&3886  &&&1897     &&&1231    &&&1001\\
$n \,(p=0.8)$	&&& 123    &&& 1074     &&&	5009  &&&	4887     &&&	3117    &&&2579\\
$\mathcal{H}$  &&& 0.0029 &&& 0.0003 &&& 0.0009 &&& 0.0007 &&& 0.0001 &&& 0.0001\\
$\mathcal{KS}$   &&& 0.0422 &&& 0.0122 &&& 0.0013 &&& 0.0047 &&& 0.0000 &&& 0.0000\\
\hline
\end{tabular}
}
\end{table}

\begin{figure}[!htbp]
\centering
\caption{Hellinger (left panel) and Kolmogorov-Smirnov (right panel) distances between $\mathcal{B}(a, 6)$ and $\mathcal{K}(\widetilde{\alpha}, \widetilde{\beta})$ distributions as a function of $a$.}
\begin{tabular}{cc}
\includegraphics[width=0.45\textwidth]{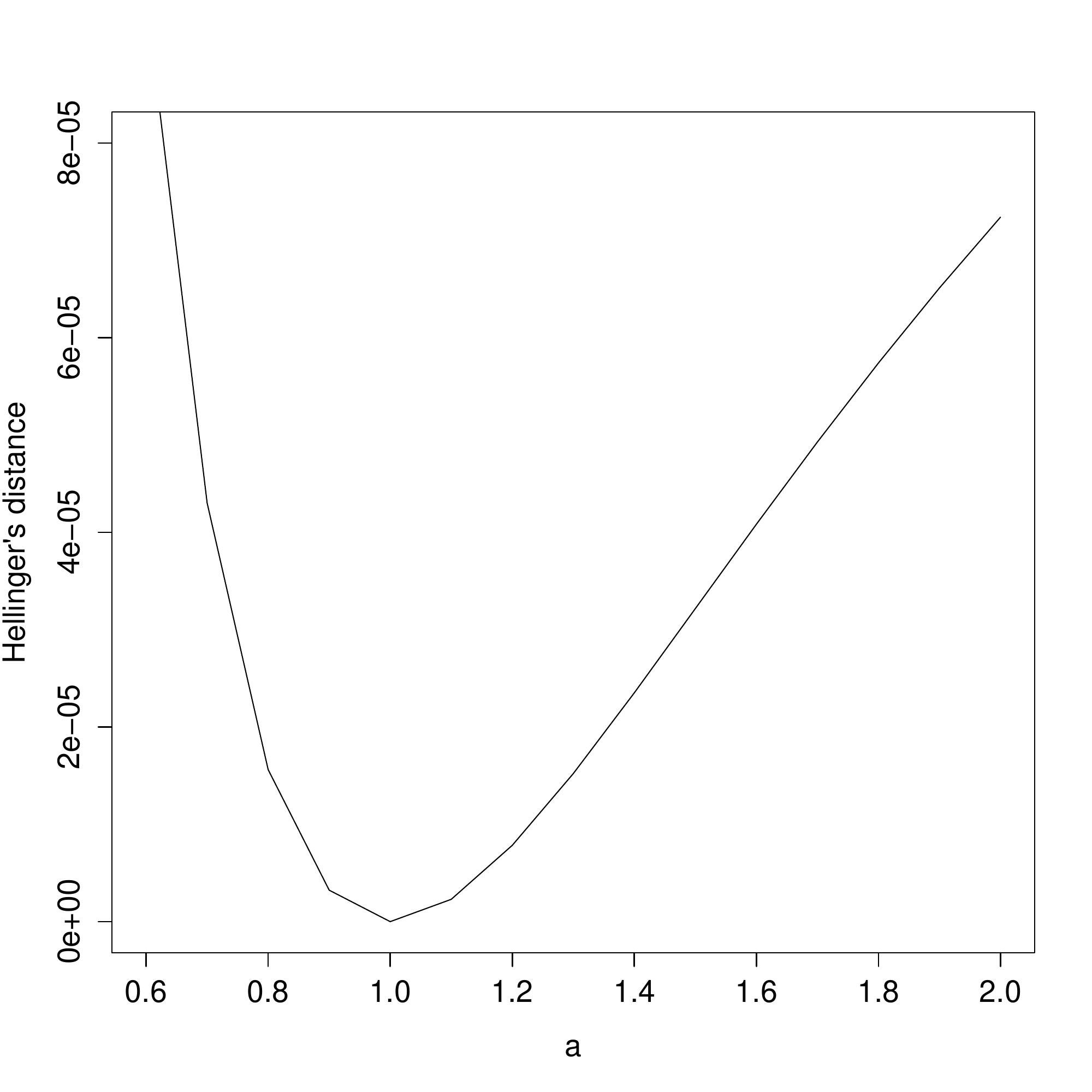}&\includegraphics[width=0.45\textwidth]{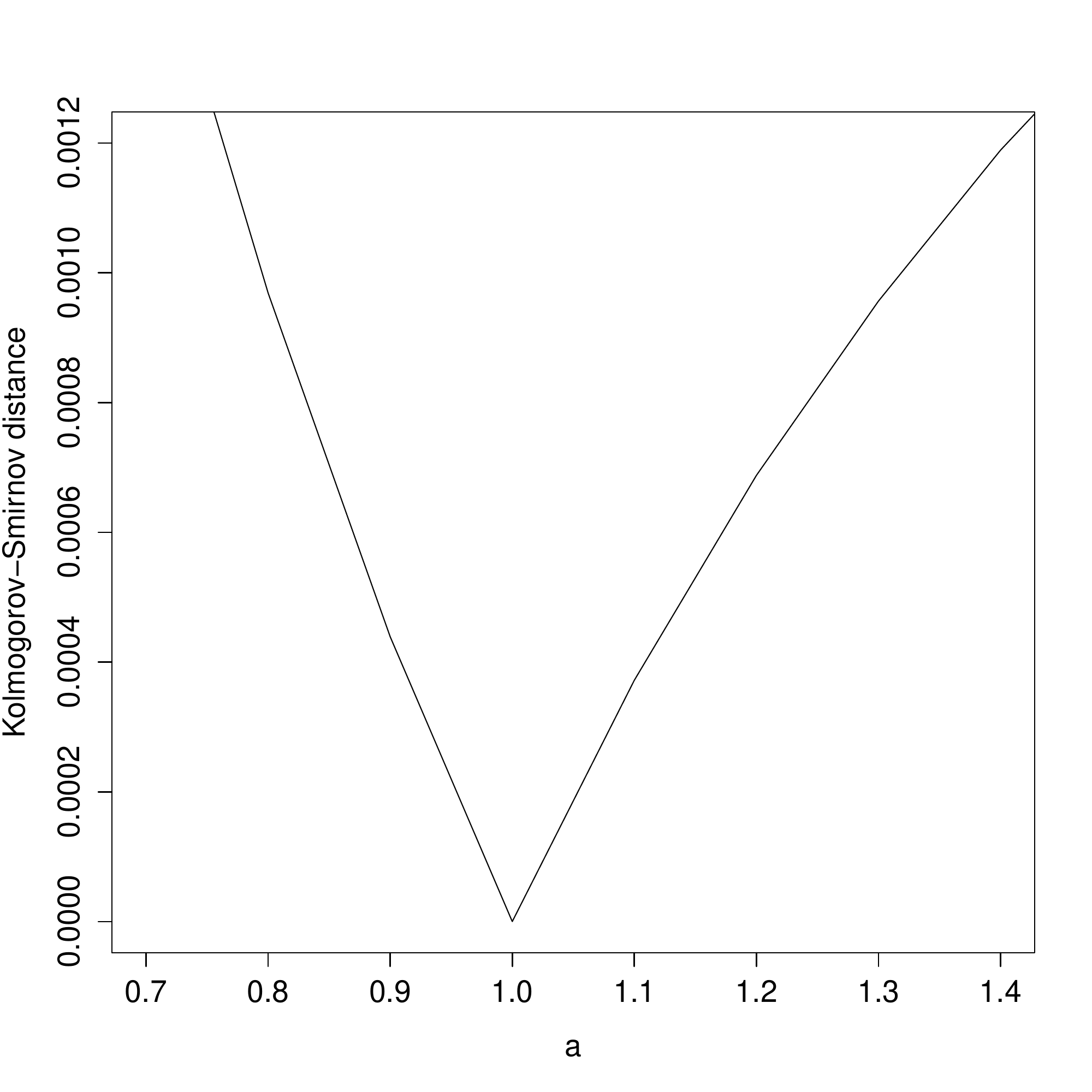}\\
\end{tabular}		
	\label{fig.Bedistances}
\end{figure}

\begin{figure}[!htbp]
\centering
\caption{Hellinger (left panel) and Kolmogorov-Smirnov (right panel) distances between $\mathcal{B}(\widetilde a, \widetilde b)$ and $\mathcal{K}(\alpha, 3)$ distributions as a function of $\alpha$.}
\begin{tabular}{cc}
\includegraphics[width=0.45\textwidth]{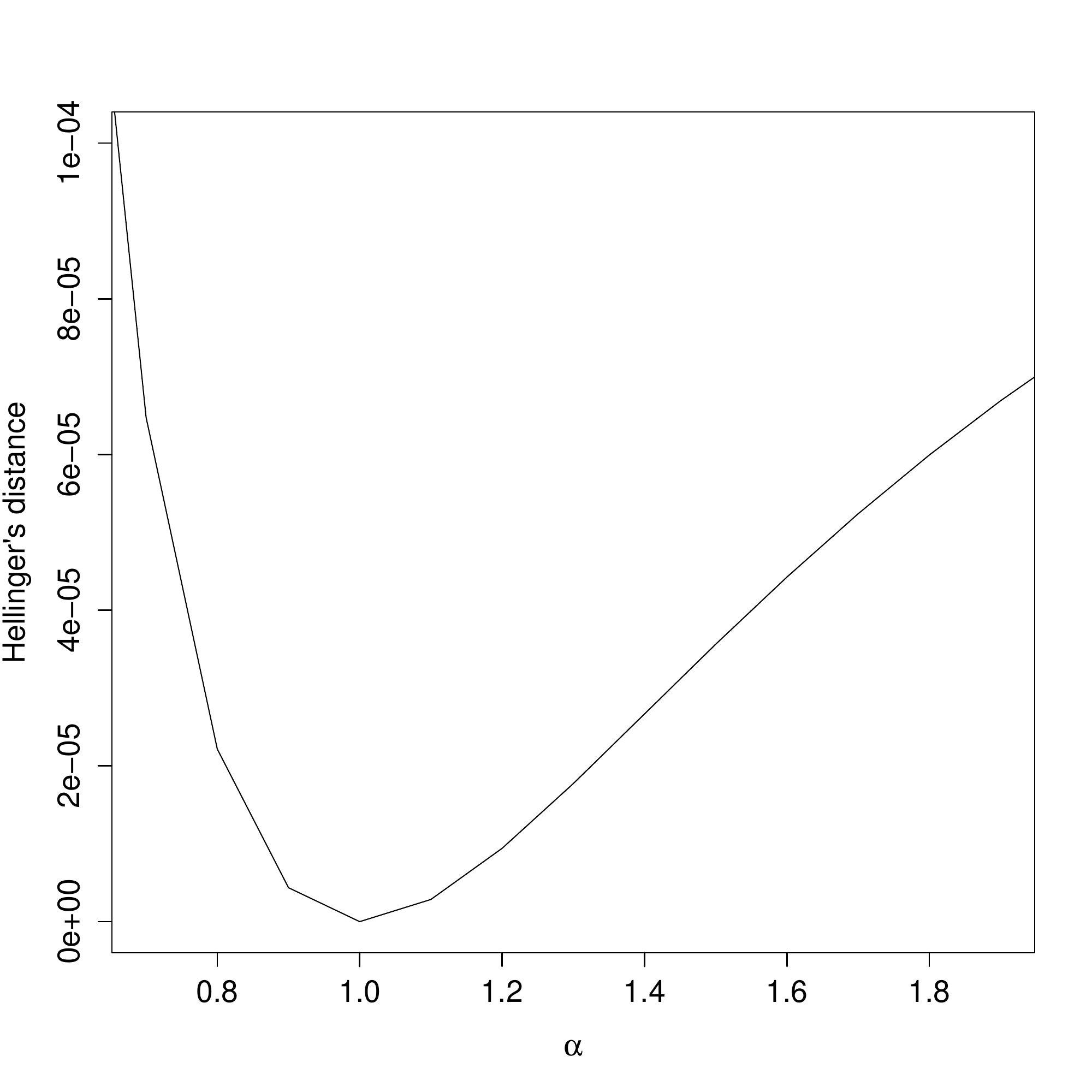}&\includegraphics[width=0.45\textwidth]{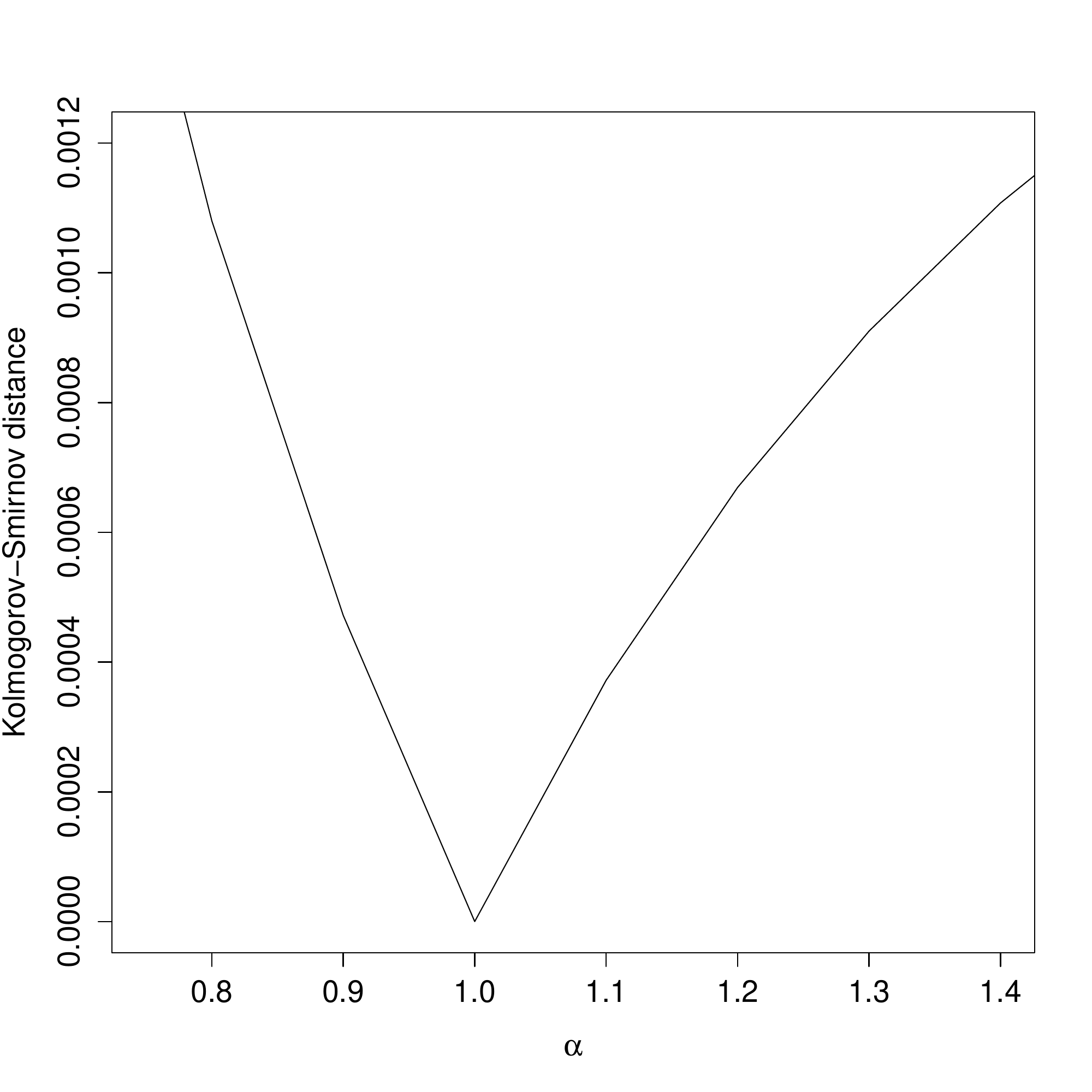}\\
\end{tabular}		
	\label{fig.Bedistances2}
\end{figure}

We shall now briefly discuss how to use the ${\rm PCS}$ and the tolerance level in a practical situation. Suppose one is interested in discriminating the beta and Kumaraswamy models where the null hypothesis is $H_{\mathcal{B}}$. Further, suppose that the tolerance level is based on the $\mathcal{H}$ distance and fixed at 0.0002. Therefore, from the Table \ref{ksbe} one needs to take the sample size $n\geq 692$ for $p=0.7$ to discriminate the beta and Kumaraswamy distributions. For a more accurate result, under the hypothesis $H_{\mathcal{B}}$ ($H_{\mathcal{K}}$), a greater range of $a$ (and $\alpha$) is required, as it is illustrated in Figure~\ref{fig.Bedistances} (Fig.~\ref{fig.Bedistances2}).    

\section{Numerical experiments}

\subsection{Simulation}\label{simulation}

We here perform some numerical experiments to observe how our asymptotic results derived in Section~\ref{Likelihoodratiotest} work for different sample sizes. We are interested in comparing the asymptotic PCSs under the hypothesis $H_{\mathcal{B}}$ and $H_{\mathcal{K}}$ with respect to the simulated probabilities based on Monte Carlo simulations.

Let us now to describe how the simulated results are obtained. We begin with the case where the null hypothesis is $H_\mathcal{B}$. The following procedure holds in a similar way for the null hypothesis $H_{\mathcal{K}}$ and therefore is omitted. Let $N$ be the number of loops of the Monte Carlo simulation and $I = (I_1, \ldots, I_N)^\top$ be a vector of length $N$. The steps, for each loop $j$, are as follows:
\begin{itemize}
\item[i)] Generate a random sample from the $\mathcal{B}$($a,b$) distribution with size $n$;
\item[ii)] Find the MLEs of $(a,b)$ and $(\alpha, \beta)$ based on the beta and Kumaraswamy distributions, respectively; 
\item[iii)] Compute the statistic $T_n = \ell^{(n)}_{\mathcal{B}}(\hat a_n, \hat b_n) - \ell^{(n)}_{\mathcal{K}}(\hat \alpha_n, \hat \beta_n)$;
\item[iv)] If $T_n >0$ take $I_j=1$, otherwise $I_j=0$.
\end{itemize}

After running the above Monte Carlo simulation, the simulated PCS is given by $\sum^N_{j=1}I_j/N$. We also compute the PCS based on the asymptotic results derived in Section \ref{Likelihoodratiotest}. The simulation study was carried out using the software \texttt{R}; see \texttt{http://www.r-project.org}.

We set $n=20, 40, 60, 80, 100, 200, 500$ and $a=0.2, 0.5, 0.9, 1.5, 2.0, 3.0, 5.0$. These results are presented in Table~\ref{simbe}. It is quite clear that there is a good agreement between the asymptotic and empirical probabilities, mainly for moderate and large sample sizes. We also observe that, when $a$ approaches 1, the PCSs approaches 0.5. This was expected since when $\alpha$ goes to 0 both beta and Kumaraswamy distributions converge to the same law. Another expected result we observed is that when $n$ increases the PCS approaches one.

In Table~\ref{simkw} we present the asymptotic and simulated PCSs under the null hypothesis $H_{\mathcal{K}}$ for $\alpha=0.2, 0.5, 0.9, 1.5, 2.0, 3.0, 5.0$ and $n=20, 40, 60, 80, 100, 200, 500$. In this case we also observe a good agreement between the asymptotic and empirical PCSs. When $\alpha$ is close to one, the PCSs are close to 0.5, and as $n$ increases, the probabilities goes to one, as expected and discussed in the previous case.  

\begin{table}[!htbp]
\centering
\caption{PCS based on the Monte Carlo simulation and based on the asymptotic results under $H_{\mathcal{B}}$ for some values of $a$ and $n$.}\label{simbe}
\scalebox{1}[0.87]{
\renewcommand{\arraystretch}{1.3}
\begin{tabular}{cccccccccccccccc}
\hline
&&  \multicolumn{13}{c}{Asymptotic probability under $H_{\mathcal{B}}$}\\
$a\downarrow n\rightarrow$ && 20 && 40 && 60 && 80 && 100 && 200 && 500\\
\hline
0.2&&	0.6669&&	0.7291&&	0.7725&&	0.8058&&	0.8326&&	0.9137&&	0.9845\\
0.5&&	0.5755&&	0.6062&&	0.6293&&	0.6484&&	0.6649&&	0.7265&&	0.8296\\
0.9&&	0.5071&&0.5100&&0.5122&&0.5141&&0.5158&&0.5223&&0.5352\\
1.5&&	0.5365&&	0.5516&&	0.5631&&	0.5727&&	0.5812&&	0.6140&&	0.6766\\
2.0&&	0.5574&&	0.5809&&	0.5988&&	0.6137&&	0.6266&&	0.6761&&	0.7649\\
3.0&&	0.5717	&&0.6009	&&0.6229	&&0.6411	&&0.6570	&&0.7162	&&0.8270	\\
5.0&&	0.5940	&&0.6254	&&0.6475	&&0.6650	&&0.6850	&&0.7500	&&0.8520	\\
\hline
&&  \multicolumn{13}{c}{Empirical probability under $H_{\mathcal{B}}$}\\ 
$a\downarrow n\rightarrow$ && 20 && 40 && 60 && 80 && 100 && 200 && 500\\
\hline
0.2&&	0.7040&&	0.7370&&	0.7890&&	0.8120&&	0.8350&&	0.9280&&	0.9840\\
0.5&&	0.5760&&	0.6090&&	0.6400&&	0.6480&&	0.6640&&	0.7200&&	0.8270\\
0.9&&   0.4934&&0.5002&&0.4980&&0.5072&&0.5040&&0.5018&&0.5260\\
1.5&&	0.5380&&	0.5400&&	0.5500&&	0.5750&&	0.5730&&	0.6280&&	0.6790\\
2.0&&	0.5900&&	0.5830&&	0.5680&&	0.5990&&	0.6090&&	0.6930&&	0.7690\\
3.0&&	0.5828	&&0.6112	&&0.6256	&&0.6438	&&0.6562	&&0.7126	&&0.8146	\\
5.0&&	0.5870	&&0.6221	&&0.6683	&&0.6799	&&0.6885	&&0.7665	&&0.8642	\\
\hline
\end{tabular}}
\end{table}

\begin{table}[!htbp]
\centering
\caption{PCS based on the Monte Carlo simulation and based on the asymptotic results under $H_{\mathcal{K}}$ for some values of $\alpha$ and $n$.}\label{simkw}
\scalebox{1}[0.87]{
\renewcommand{\arraystretch}{1.3}
\begin{tabular}{cccccccccccccccc}
\hline
&&  \multicolumn{13}{c}{Asymptotic probability under $H_{\mathcal{K}}$}\\ 
$\alpha\downarrow n\rightarrow$ && 20 && 40 && 60 && 80 && 100 && 200 && 500\\
\hline
0.2&&	0.7778&&	0.8602&&	0.9073&&	0.9369&&	0.9563&&	0.9922&&	0.9999\\
0.5&&	0.6458&&0.6645&&	0.6788&&0.6908&&	0.7013&&	0.7418&&	0.8171\\
0.9&&	0.5053&&0.5074&&0.5091&&0.5105&&0.5118&&0.5166&&0.5263\\
1.5&&	0.5266&&0.5976&&0.6161&&0.6632&&0.7194&&0.7536&&0.7908\\
2.0&&	0.6059&&	0.6383&&	0.6802&&	0.7518&&	0.7931&&	0.8186&&	0.8594\\
3.0&&	0.5944&&	0.6162&&	0.6877&&	0.7788&&	0.8599&& 0.9039&&0.9520\\
5.0&&	0.6295&&0.6417&&0.6511&&0.6589&&0.6658&&0.6926&&0.7445\\
\hline
&&  \multicolumn{13}{c}{Empirical probability under $H_{\mathcal{K}}$}\\ 
$\alpha\downarrow n\rightarrow$ && 20 && 40 && 60 && 80 && 100 && 200 && 500\\
\hline
0.2&&	0.8250&&	0.8400&&	0.8970&&	0.9220&&	0.9520&&	0.9930&&	0.9990\\
0.5&&	0.6360&&0.6548&&0.6654&&0.6894&&0.7038&&0.7406&&	0.8116\\
0.9&&	0.5048&&0.5246&&0.5050&&0.5190&&0.5254&&0.5264&&0.5332\\
1.5&&	0.4624&&0.5866&&0.6104&&0.6682&&0.7088&&0.7590&&0.7824\\
2.0&&	0.6060&&	0.6240&&	0.6870&&	0.7280&&	0.7490&&	0.8130&&	0.8760\\
3.0&&	0.5950&&	0.6380&&	0.6700&&	0.7700&&	0.8600&&	0.8900&&	0.9330\\
5.0&&	0.5592&&	0.5880&&0.6120&&0.6204&&0.6224&&0.6658&&0.7272\\
\hline
\end{tabular}}
\end{table}

\subsection{Empirical illustrations}
\label{applications}

We now apply our results in two real data sets. In the first application, we consider the percentage of muslim population in 152 countries. The data can be found in \texttt{http://www.qran.org/a/a-world.htm} and is based on 2004 Census projection. The sources include \texttt{HFE.org}, \texttt{IslamicPopulation.com}, \texttt{StrategicNetwork.org}, \texttt{State.gov}, among others.

\begin{figure}[!htbp]
\centering
\caption{Estimated densities (left panel) and cumulative (right panel) functions for the first data set.} \label{muslim}
\begin{tabular}{cc}
\includegraphics[width=0.5\textwidth]{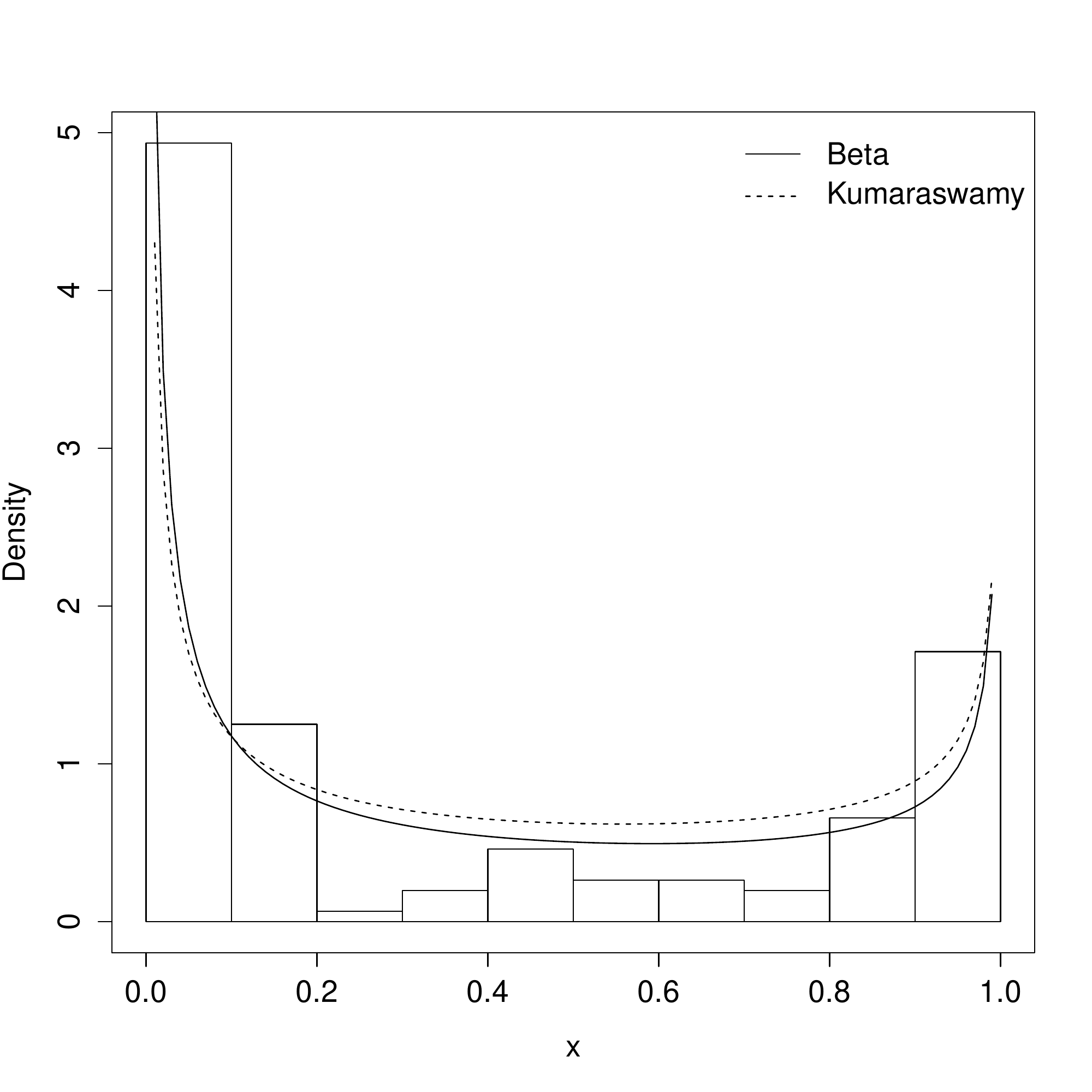}&\hskip-5mm\includegraphics[width=0.5\textwidth]{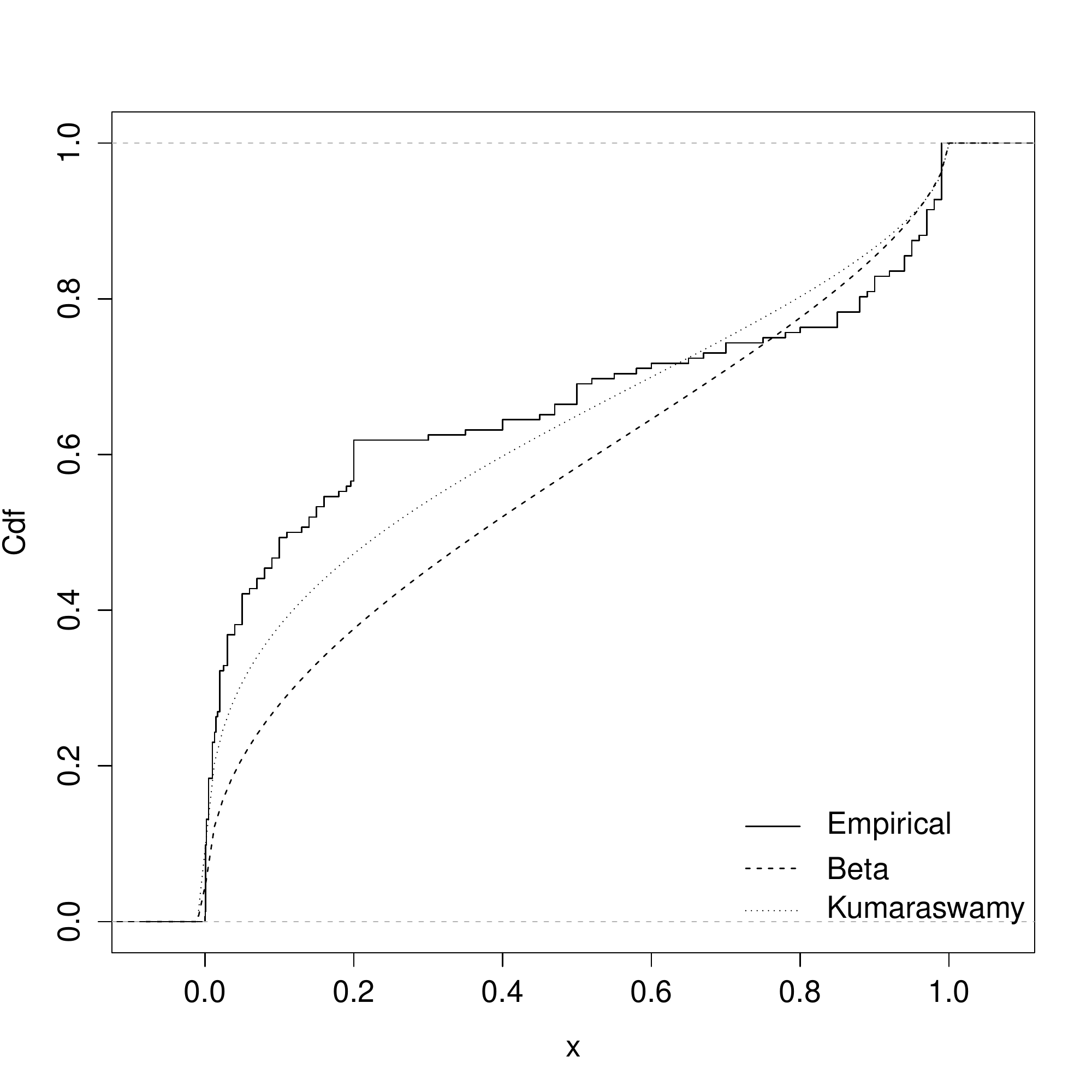}
\end{tabular}		
\end{figure}

The MLEs of the parameters of the beta and Kumaraswamy distributions are given by $(\widehat a, \widehat b) = (0.2976, 0.5159)$ and $(\widehat \alpha, \widehat \beta) = (0.3515, 0.5906)$, respectively. Figure~\ref{muslim} shows the histogram and the plots of the fitted beta and Kumaraswamy densities. Empirical and fitted cdfs are also displayed in this figure.  
The test statistic equals $T_n = 118.4542-114.8334 = 3.6207 > 0$, which indicates that the beta model should be chosen according Akaike criterion. Under the hypothesis that the data come from a $\mathcal B(0.2976, 0.5159)$ distribution, we obtain the estimated quantities $\mbox{AM}_{\mathcal{B}}(\widehat a, \widehat b) = -0.0188$ and $\mbox{AV}_{\mathcal{B}}(\widehat a, \widehat b) = 0.1617$. Thus, we have $\mathrm{PCS_{\mathcal{B}}}(\widehat a, \widehat b) = 0.7174$, while the simulated PCS equals $0.7370$. Similarly, under the hypothesis that the data come from a Kumaraswamy distribution, we have $\mbox{AM}_{\mathcal{K}}(\widehat \alpha, \widehat \beta) = -0.0026$ and $\mbox{AV}_{\mathcal{K}}(\widehat \alpha, \widehat \beta) = 0.0380$, which yields $\mathrm{PCS_{\mathcal{K}}}(\widehat \alpha, \widehat \beta) = 0.5917$ (the simulated PCS equals 0.6180). Therefore, the probability of correct selection (based on the asymptotic result) is at least equal to $\min\left\{0.7174, 0.5917\right\} = 0.5917$. Since the PCS is maximum under the hypothesis $H_{\mathcal{B}}$, we choose the beta distribution. Based on the simulated PCSs, we obtain the same conclusion.\\

The second application considers the proportion of atheists in the populations of 137 countries. This data set was also used by Lynn et al. (2009) and collected from surveys mostly carried out in 2004, although in a few countries the surveys were a year or two earlier.

\begin{figure}[!htbp]
\centering
\caption{Estimated densities (left panel) and cumulative (right panel) functions for the second data set.}
\begin{tabular}{cc}
\includegraphics[width=0.5\textwidth]{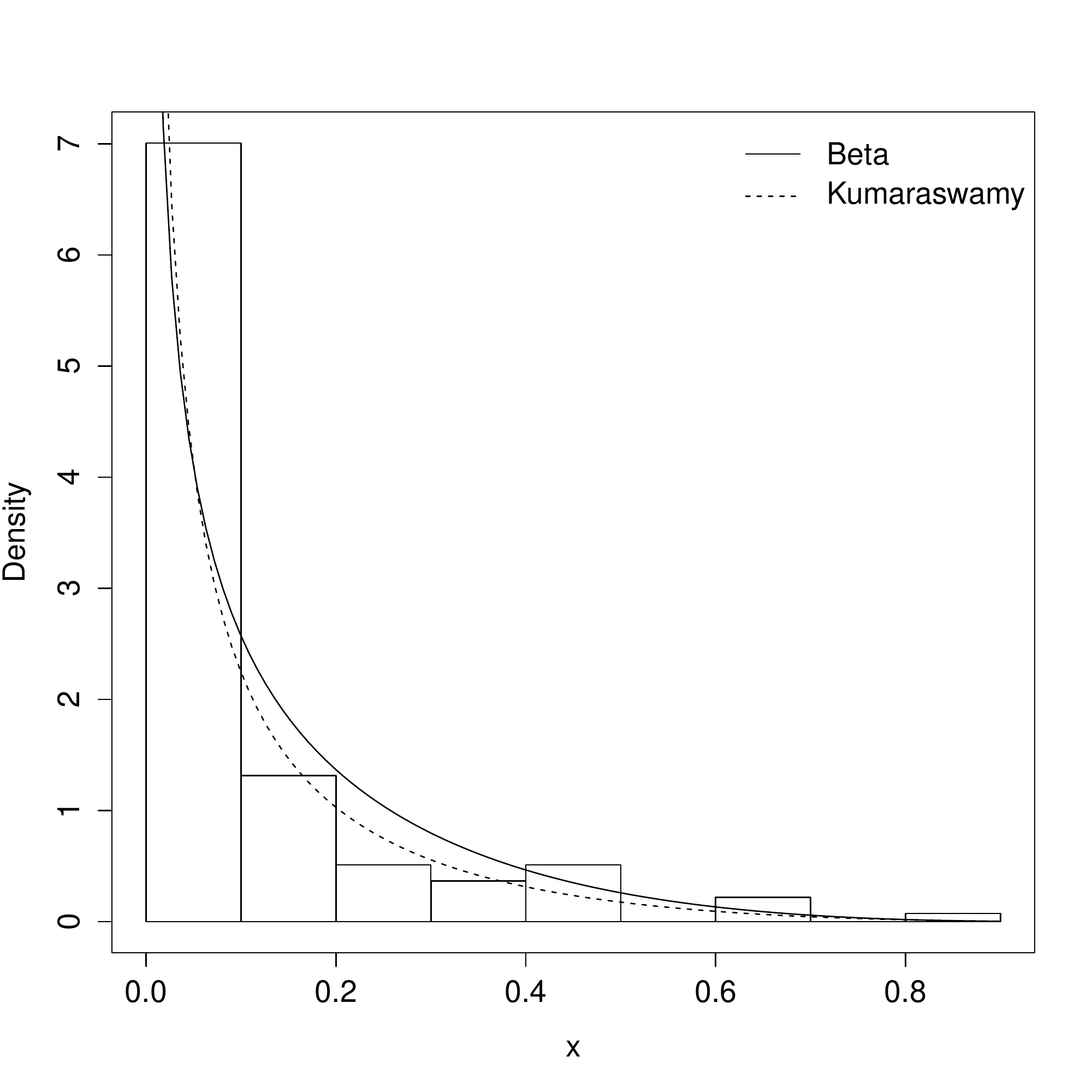}&\hskip-5mm\includegraphics[width=0.5\textwidth]{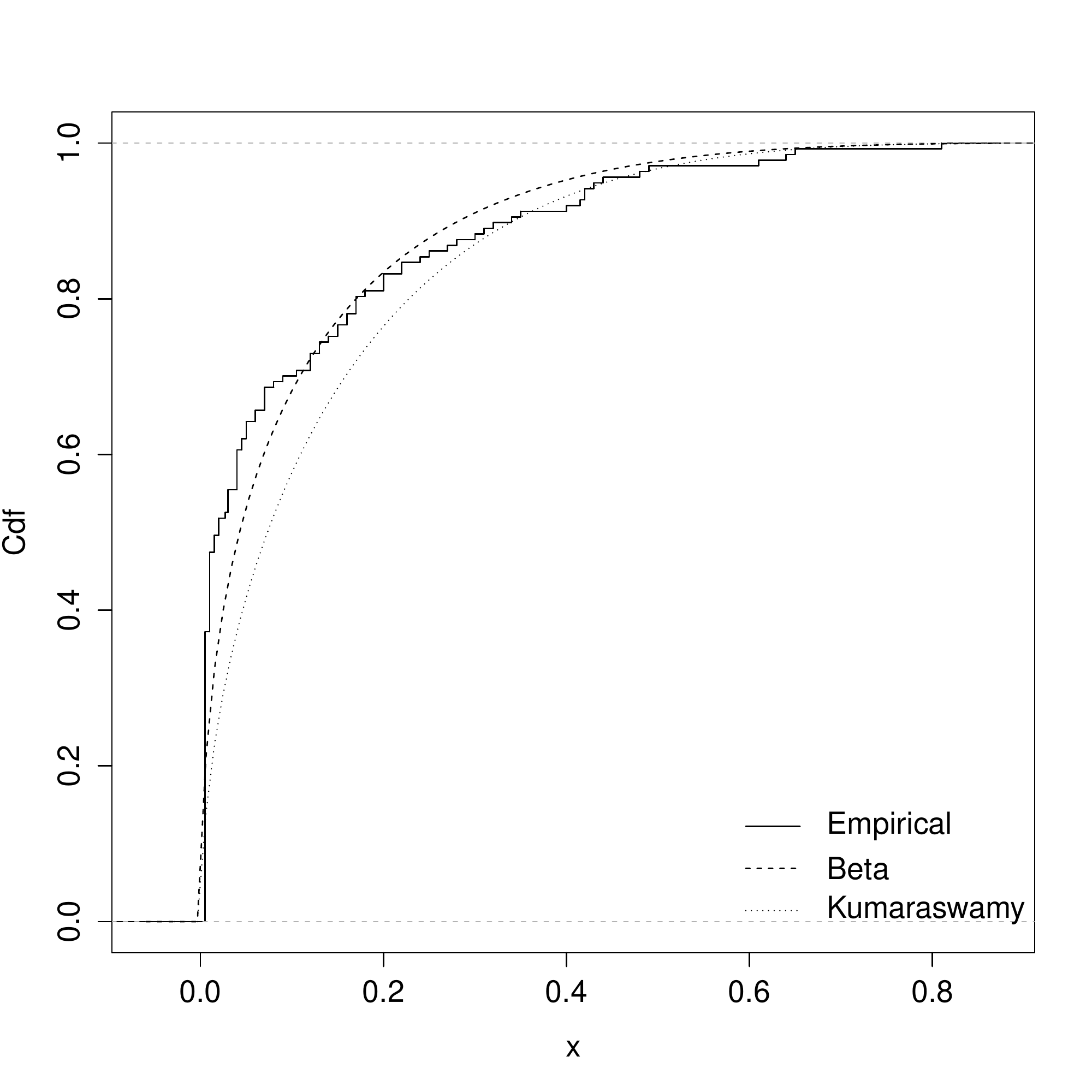}\\
\end{tabular}		
	\label{fig.estimation4}
\end{figure}

 The MLEs of the beta and Kumaraswamy parameters are $(\widehat a, \widehat b) = (0.4368, 3.6347)$ and $(\widehat \alpha, \widehat \beta) = (0.5091, 3.0914)$. The histogram of the data and the beta and Kumaraswamy estimated densities are shown in Figure~\ref{fig.estimation4}. For comparison purposes, we also plot empirical and the two fitted cdfs. In this case, the test statistic equals $T_n = 205.9754 - 210.8923 = -4.9169<0$, thus indicating that the Kumaraswamy model yields the best fit (based on the Akaike criterion). Under the hypothesis $H_\mathcal K$, we have $\mbox{AM}_{\mathcal{K}}(\widehat \alpha, \widehat \beta) = 0.0035$ and $\mbox{AV}_{\mathcal{K}}(\widehat \alpha, \widehat \beta) = 0.0072$, and hence we obtain $\mathrm{PCS_{\mathcal{K}}}(\widehat \alpha, \widehat \beta) = 0.7872$; the simulated PCS equals $0.8044$. On the other hand, under the hypothesis $H_\mathcal B$, we obtain $\mbox{AM}_{\mathcal{B}}(\widehat a, \widehat b) = -0.0032$ and $\mbox{AV}_{\mathcal{B}}(\widehat a, \widehat b) = 0.0063$. With these results we find $\mathrm{PCS_{\mathcal{B}}}(\widehat a, \widehat b) = 0.6812$ and the simulated PCS equals $0.7060$. The probability of correct selection (based on the asymptotic results) is at least $\min(0.7872, 0.6812) = 0.6812$. The PCS is maximum under the hypothesis $H_{\mathcal{K}}$ and therefore we choose the Kumaraswamy distribution. The same conclusion is obtained by considering the simulated results.\\

\section*{Acknowledgements}
The authors gratefully acknowledge financial support from CAPES (Brazil) and CNPq (Brazil).

\end{document}